\documentclass[final]{agujournal2019}
\usepackage{url} 
\usepackage{lineno}
\usepackage{amsmath}
\usepackage{amsfonts}
\usepackage{float}

\draftfalse

\journalname{JGR: Planets}

\begin{document}

\let\oldcite\cite
\renewcommand{\cite}[1]{\textcolor{blue}{\oldcite{#1}}}
\let\oldciteA\citeA
\renewcommand{\citeA}[1]{\textcolor{blue}{\oldciteA{#1}}}

%
%


\title{Inferring Fireball Velocity Profiles and Characteristic Parameters of Meteoroids from Incomplete Datasets}

\authors{Eloy Peña-Asensio\affil{1} and Maria Gritsevich\affil{2,3,4}}

\affiliation{1}{Department of Aerospace Science and Technology, Politecnico di Milano, Via La Masa 34, 20156 Milano, Italy}
\affiliation{2}{Swedish Institute of Space Physics (IRF), Bengt Hultqvists väg 1, 981 92 Kiruna, Sweden}
\affiliation{3}{Faculty of Science, Gustav H\"{a}llstr\"{o}min katu 2, FI-00014 University of Helsinki, Finland}
\affiliation{4}{Institute of Physics and Technology, Ural Federal University, Mira str. 19, 620002 Ekaterinburg}

\correspondingauthor{Eloy Peña-Asensio}{eloy.pena@polimi.it eloy.peas@gmail.com}

\begin{keypoints}
    \item Retrieving fireball parameters from partial data with global metaheuristic optimization algorithms.
    \item Delimitation of fireball velocity profiles using only the beginning and terminal points.
    \item Reconciliation of photometric measurements and a purely dynamical model for asteroidal meteoroids from EN catalog.
\end{keypoints}

\begin{abstract}

Extracting additional information from old or incomplete fireball datasets remains a challenge. To address missing point-by-point observations, we introduce a method for estimating atmospheric flight parameters of meteoroids using metaheuristic optimization techniques. Using a fireball catalog from the European Fireball Network (EN), we reconstruct velocity profiles, meteoroid bulk densities, mass loss rates, and ablation and ballistic coefficients, based on the initial and terminal points' height, velocity, and mass with the purely dynamical $\alpha$-$\beta$ model. Additionally, the method's performance is compared to the Meteorite Observation and Recovery Project (MORP) derived fits, confirming the robustness of the computed parameters for objects with asteroidal compositions. Our findings show that $\alpha$-$\beta$ model yields parameters consistent with the photometric and dynamic mass estimates in the EN catalog for $P_E$ type I events. However, in the implementation proposed here, $\alpha$-$\beta$ model encounters limitations in accurately representing the final deceleration of more fragile high-velocity meteoroids. This is likely due to challenges in representing complex fragmentation processes by fitting only two points, even when initial and terminal residuals are minimal. The retrieved $\alpha$-$\beta$ distribution differs from the one derived from MORP data, likely due to the imposed mass constraints, which strongly influence the results, especially the bulk density. The results suggest that $P_E$ constraints reduce fitting accuracy (from 90\% to 44\%), while flexibility and freedom from assumptions improve $\alpha$-$\beta$ performance. The method yields 26\% of the events compatible with the catalog $P_E$ classification. Our approach is well-suited for interpreting historical or sparse datasets.

\end{abstract}

\section*{Plain Language Summary}

This study offers a new way to analyze fireballs—bright shooting stars produced by the impact of space rocks on the atmosphere—using advanced computer tools. As sometimes the full observational data of fireballs is not available, our method helps fill in the gaps by estimating their speed, composition, and how they disintegrate in the Earth's atmosphere. We focused on fireballs recorded by the European Fireball Network. Our results were particularly accurate for slow rocks similar to asteroids but showed some challenges with weaker and faster ones (e.g., those of cometary origin) or those previously flagged as iron. The mass loss and aerodynamic characteristics we obtain are different than expected, possibly because, unlike previous studies, we have taken into account the initial and final masses of the objects, which influence the results. We also found that using certain assumptions based on the fireball type could make our modeling less accurate. By comparing our estimates with a complete dataset, we confirmed the reliability of our approach. This new method can help us to analyze incomplete fireball data.

\section{Introduction} \label{sec:intro}


Meteoroids, originating from asteroidal, cometary, or planetary bodies, traverse interplanetary space and range in size from 30 $\mu$m to 1 m in diameter \cite{Koschny2017JIMO}. When small meteoroids enter planetary atmospheres at enormous velocities, they generate the luminous phenomena known as meteors due to ablation caused by inelastic collisions with air molecules \cite{Ceplecha1998SSRv}. For larger fireball-class meteoroids (centimeter-sized or larger), the primary mechanism is compressive heating caused by high ram pressure generated by their hypersonic speeds \cite{Silber2018AdSpR}. This pressure induces significant mechanical stress, leading to fragmentation when it exceeds the tensile strength of the meteoroid material. As the meteoroid fragments, fresh surfaces are exposed to intense ablation, further increasing the mass loss rate. The combined effects of ablation and fragmentation result in the bright fireballs and potential airbursts observed during these events.


Fireball trajectory and velocity determinations require multistation detection, as single-station observations do not usually provide precise positional data. While optical cameras provide substantial observational video data, their resolution frequently falls short of accurately capturing deceleration profiles. As a result, traditional mass estimates are often based on empirical relations or light curves. This constraint has spurred extensive debates over dynamic versus photometric mass estimations, which were eventually clarified by matching suitable bulk density values and recognizing the critical role of fragmentation \cite{Ceplecha1998SSRv, Ceplecha2005, Popova2019msmebook, Borovicka2019msmebook}. However, some studies integrate both dynamical and photometric data along with fragmentation models, aiming to reconstruct the entry trajectory in finer detail. Despite increasing the number of free parameters, recent approaches demonstrate the effectiveness of this combined methodology in providing a more comprehensive representation of meteoroid mass evolution by simultaneous fitting the deceleration and emitted light curve \cite{Borovicka2020AJ, VIDA2024115842}.



There has been growing interest in recent years in exploring new approaches to infer fireball parameters. These methods leverage advancements in computational techniques, data science, and theoretical modeling to extract valuable information from diverse observational datasets \cite{Johnston2018Icar, Colonna2019hmepbook, Tarano2019Icar, Register2020Icar, Hulfeld2021AA, Brykina2023FlDy, Li2023AMS, VIDA2024115842}. This tendency represents an effort to harness the potential of contemporary knowledge for enhanced meteoroid characterization. With this work, our goal is to tackle the challenges of analyzing databases lacking point-by-point observations. In doing so, we pave the way for a deeper understanding of meteoroid properties, as we aim to bolster the development of robust and versatile tools for characterizing fireballs. These tools will facilitate automated classification of meteor phenomena, thereby improving our ability to promptly recover unweathered meteorites as well as effectively monitor and mitigate potential threats from near-Earth objects.

In Section \ref{sec:model_class}, we revisit and refine the dynamic approach involving luminous trajectory reconstruction, which employs a neat dimensionless parameterization to estimate the mass of a meteoroid and its variation over time when penetrating the atmosphere. To be applicable, this approach generally requires knowledge of height and velocity at a minimum of three points along the trajectory, one of which coincides with the entry velocity \cite{Gritsevich2009AdSpR, Lyytinen2016PSS}. To overcome this requirement, Section \ref{sec:inferring} introduces a novel method for inferring fireball parameters that utilizes two trajectory points and information derived from photometric observations, bypassing the need for direct deceleration curves. Inspired by reverse engineering processes, we explore global optimization algorithms to reconstruct the atmospheric flight of meteoroids from incomplete trajectory information employing their total energy measurements and pure atmospheric dynamic models. We apply this method to the European Fireball Network (EN) data catalog \cite{Borovicka2022AA_I, Borovicka2022AA_II} and present the results in Section \ref{sec:results}. Additionally, we evaluate the retrieved parameters by assessing their efficacy in predicting terminal height through various models and by comparing these predictions with a database where complete velocity profiles were utilized \cite{Gritsevich2008DokPh, Moreno2015Icar}. We also compare our reconstructed velocity profiles with the only five available to us cataloged examples containing complete point-by-point measurement data. In Section \ref{sec:conclusions}, we summarize our conclusions.


\section{Theoretical background of meteoroid entry} \label{sec:model_class}

\subsection{Equations of motion}

The equations governing the motion of a meteoroid, see, e.g.,  \citeA{gritsevich2017constraining}, can be expressed as follows:

\begin{equation}
M\frac{dV}{dt}=-\frac{1}{2}c_{d}\rho V^{2}S, \label{1}
\end{equation}

\begin{equation}
\frac{dh}{dt}=-V\sin \gamma, \label{2}
\end{equation}

\begin{equation}
H^{\ast }\frac{dM}{dt}=-\frac{1}{2}c_{h}\rho V^{3}S. \label{3}
\end{equation}


These equations can be employed to describe the motion of a meteoroid in planetary atmospheres \cite{CHRISTOU2024116116}, and they are defined concerning various physical quantities, as summarized in \ref{sec:defs}. There are numerous methods available for handling these equations to obtain practical results, many of which entail making assumptions about the unknowns to compute the variables of interest. We adopt the method of non-dimensionalization, initiated through dimensional analysis, due to its effectiveness in eliminating or reducing the dimensions of equations. This technique scales quantities by characteristic units or physical constants, providing insights into the fundamental properties of the system, as demonstrated in prior studies \cite{Stulov1997ApMRv, Gritsevich2006SoSyR}.

The derivation of the model involves computing sets of dimensionless parameters from given variables, known as non-dimensionalization within the Buckingham $\pi$ theorem (\ref{sec:dim}). Careful selection of fundamental quantities is crucial to ensure the independence of their dimensions when solving physics problems using dimensional analysis. In our case, time can be eliminated using Eq. \ref{2}, leaving the independent variables as $M$, $V$, and $h$, and the dependent variables as $\rho$ (a function of height) and $S$, which can be expressed through $M$. Nondimensionalization involves replacing each variable with a scaled quantity relative to a characteristic unit of measure, chosen judiciously to simplify the equations, see \ref{sec:defs}, where all notations are given.


The resulting equations describing changes in normalized mass and velocity with height in new dimensionless variables can be written as \cite{Gritsevich2007SoSyR}:

\begin{equation}
m\frac{dv}{dy}=\frac{1}{2}c_{d}\frac{\rho_{sl}h_0S_{beg}}{M_{beg}}\frac{\rho vs}{\sin \gamma }, \label{4a}
\end{equation}

\begin{equation}
\quad \frac{dm}{dy}=\frac{1}{2}c_{h}\frac{\rho_{sl}h_0S_{beg}}{M_{beg}}\frac{V_{beg}^{2}}{H^{\ast }}\frac{\rho v^2s}{\sin \gamma }. \label{4b}
\end{equation}

Under the assumption $s=m^{\mu }$, an isothermal atmosphere ($\rho =e^{-y}$), and with the boundary conditions: $y=\infty$, $v=1$, $m=1$ equations (\ref{4a}) and (\ref{4b}) have the analytical solution:

\begin{equation}
m(\beta,\mu,v)= e^{-\beta \frac{1-v^{2}}{1-\mu }}\label{6_mass}
\end{equation}
and
\begin{equation}
y(\alpha,\beta,v)=\ln 2\alpha +\beta -\ln (\overline{E}i(\beta )-\overline{E}i(\beta v^{2})), \label{eq_y}
\end{equation}

where 
\[
\overline{Ei}(x)=\int_{-\infty }^{x}\frac{e^{z}dz}{z} 
\]

is the exponential integral (\ref{sec:exp_int}) and the ballistic coefficient $\alpha$ and the mass loss parameter $\beta$ were defined as:

\begin{equation}
\alpha =\frac{c_{d}\rho_{sl}h_0S_{beg}}{2M_{beg}\sin \gamma }, \label{a}
\end{equation}

\begin{equation}
\beta =\frac{(1-\mu )c_{h}V_{beg}^{2}}{2c_{d}H^{\ast }}. \label{b}
\end{equation}

By using the definition of the ablation coefficient:

\begin{equation}
\sigma =\frac{c_h}{c_d H^*}, \label{eq_abla_coeff}
\end{equation}

together with Eq. \ref{b} yields to:

\begin{equation}
\beta =\frac{1}{2}\sigma V_{beg}^2(1-\mu). \label{eq_beta_abla}
\end{equation}

Subsequently, this method reduces all the unknowns in fireball flight modeling to two key dimensionless parameters, $\alpha$ and $\beta$, which carry straightforward physical interpretations and can be uniquely determined for each event based on observations. Essentially, it categorizes fireballs into similar-consequence events by assigning specific $\alpha$ and $\beta$ values to each ``category''. 

The ballistic coefficient $\alpha$ is directly proportional to the mass of the atmospheric column along the trajectory, as encountered by the body upon entry, with the cross-section $S_{beg}$, divided by the meteoroid’s pre-atmospheric mass. Essentially, $\alpha$ quantifies the height and intensity of the drag experienced by the meteoroid during its flight. The mass loss parameter $\beta$ describes the rate at which the meteoroid sheds mass. It can be interpreted as the ratio of the initial kinetic energy possessed by the meteoroid upon entering the atmosphere to the energy required to completely destroy this object within the atmosphere.

In processing modern fireball network records this approach is particularly useful in prompt identification of meteorite-producing fireballs, demonstrating their clustering on the $\alpha$-$\beta$ plot associated with varying probabilities of resulting in meteorites on the ground, with meteoroids having different likelihoods of producing meteorites. Indeed, the determined values of $\alpha$ and $\beta$ for observed meteorite falls, such as Pribram, Lost City, Innisfree, Neuschwanstein \cite{Gritsevich2008SoSyR}, Benešov \cite{Gritsevich2007SoSyR, Gritsevich2008Parameters}, Park Forest \cite{meier2017park}, Annama \cite{Trigo2015MNRAS, Lyytinen2016PSS},  Bunburra Rockhole \cite{sansom2015}, Košice \cite{gritsevich2017constraining}, Dingle Dell \cite{Devillepoix2018mps}, Murrili \cite{Sansom2020mps}, Cavezzo \cite{Gardiol2021MNRAS, Boaca2022ApJ}, Arpu Kuilpu \cite{Shober2022MAPS}, Madura Cave \cite{Devillepoix2022MPS}, Ådalen \cite{Kyrylenko2023ApJ}, Traspena \cite{Andrade2023MNRAS}, Ischgl \cite{Gritsevich2024}, and Ozerki \cite{Macente2025}, enable accurate identification of these meteorite droppers based on trajectory through this methodology. Furthermore, $\alpha$-$\beta$ characterization has been employed in many other meteor and fireball analyses \cite{Popelenskaya2008SoSyR, Bouquet2014PSS, Vaubaillon2015EMP, Eloy2021MNRAS, Vierinen2022FrASS, Eloy2023, Eloy2024Icar, Eloy2024MNRAS_IberianSuperbolide}.

It has been demonstrated that the values of $\alpha$ and $\beta$ can be employed to deduce the degree of penetration of fireballs into the atmosphere by estimating the terminal heights of their luminous flight \cite{Gritsevich2008DokPh, Moreno2015Icar, Moreno2017ASSP, Sansom2019ApJ}. The interpretation of fireball light curves is also facilitated by the known $\alpha$ and $\beta$ values, enabling a meaningful physical interpretation of the observed phenomena, allowing for inferences regarding the values of the shape change coefficient $\mu$ and luminous efficiency $\tau$ \cite{Gritsevich2011Icar, Bouquet2014PSS, Drolshagen2021AAa, Drolshagen2021AAb}.




\subsection{Mass estimation}

When interpreting meteor phenomena, dynamic parameters typically relate to the main leading fragment, unless stated otherwise and/or a suite of individual fragment trajectories is considered \cite{Moilanen2021LPICo26096288M}. Therefore, calculating the mass of the main body along the luminous path requires knowledge of $\beta$ (or $\sigma$) and $\mu$, typically ranging between 0 -no rotation- and 2/3 -isotropic surface ablation- \cite{Gritsevich2006SoSyR, Bouquet2014PSS, Sansom2019ApJ}. In the classical inverse problem approach, which involves identifying the most suitable fit to observational data, all aspects of mass loss, including fragmentation, are integrated into the model. In other words, the mass of the main fragment decreases exponentially as described by Eq. \ref{6_mass}. Bayesian filtering and Monte Carlo techniques can also be employed for mass estimation, including masses of individual fragments \cite{Sansom2017, Moilanen2021MNRAS, Gritsevich2024}. Moreover, all these estimates may vary with changes in assumptions about the atmospheric model \cite{Lyytinen2016PSS}. Traditionally, photometry can be used to estimate meteoroid mass by analyzing the fraction of kinetic energy transformed into the light emitted by the fireball \cite{McCrosky1968}. This method often assumes constant velocity and relies on poorly determined luminous efficiency coefficients, rendering it less reliable, particularly for meteorite-dropping events \cite{Ceplecha1996physics, Gritsevich2008Validity, Gritsevich2011Icar}. 

Continuing our dynamic approach and following \citeA{Gritsevich2009AdSpR}, the meteoroid mass at the beginning of the luminous trajectory can be expressed from Eq. \ref{a}: 

\begin{equation}
M_{beg}=\left(\frac{1}{2} \frac{c_d A_{beg} \rho_{sl} h_0}{\alpha \rho_m^{2/3} \sin\gamma}\right)^{3}.
	\label{eq:mass_init}
\end{equation}

Previous studies have shown that pre-atmospheric mass estimates obtained this way correlate with pre-atmospheric sizes derived from laboratory analyses of recovered meteorites using cosmogenic radionuclide analysis \cite{gritsevich2008estimating, gritsevich2017constraining, kohout2017annama, meier2017park, Gritsevich2024}, even though this approach reflects deceleration of the main fragment and does not explicitly account for fragmentation.

The terminal mass can be estimated by substituting the last observed velocity and the computed pre-atmospheric mass in Eq. \ref{6_mass}:

\begin{equation}
M_{ter}=M_{beg} e^{-\frac{\beta}{1-\mu} \left(1-\left(\frac{V_{ter}}{V_{beg}}\right)^2 \right)}.
	\label{eq:mass_terminal}
\end{equation}

This estimate can be further refined by considering the minor ablation that also occurs below the terminal point of a fireball \cite{Moilanen2021MNRAS}.

\subsection{Fireball classification}




Classifying fireballs is essential for understanding the physical properties, origin, and behavior of meteoroids as they enter the Earth's atmosphere, as well as for quickly assessing expected outcomes for their impacts. One common method for classifying fireballs is the $P_E$ criterion of \citeA{Ceplecha1976JGR}, which is based on the penetration ability of meteoroids. It is an empirical relationship aimed at helping to differentiate between meteoroid types considering the total photometric measurements. It is computed as:

\begin{equation}
P_E = \log \rho_{ter} - 0.42 \log M_{phot} + 1.49 \log V_{beg} - 1.29 \log \cos z,
\label{eq:PE}
\end{equation}

where \( \rho_{ter} \) denotes the atmospheric density at the fireball terminal height in $g\,cm^{-3}$, \( M_{phot} \) is the photometric mass in grams calculated using the original luminous efficiency \( \tau \) \cite{Ceplecha1976JGR}, \( V_{beg} \) represents the entry velocity in $km\,s^{-1}$, and \( z \) is the zenith distance of the apparent radiant. As the equation demonstrates, the $P_E$ criterion does not require any assumptions about the ablation coefficient or meteoroid bulk density. It uses physical parameters that can be deduced from observations, such as atmospheric density at the end of the trajectory, entry velocity, trajectory slope, and mass estimates based on the light curve. It also becomes apparent that for the criterion to be applicable, both the initial and terminal points of the trajectory need to be observed.

McCrosky and Ceplecha applied the $P_E$ criterion to classify 232 PN fireballs into four groups based on their $P_E$ values. After this classification, they suggested the same bulk densities and ablation coefficients for each event in these suggested groups, providing an intuitive understanding of meteoroid behavior during atmospheric entry. The four fireball types, I, II, IIIA, and IIIB (from the strongest to the weakest), were defined, with the boundary $P_E$ values being -4.60, -5.25, and -5.70. Subsequently, it was suggested that by computing the $P_E$ value allows any meteor event to be classified into one of the predefined groups, enabling the assignment of default values for the ablation coefficient and meteoroid bulk density associated with that group to the event.

\citeA{moreno2020physically} suggested that the combination of $\alpha$ and $\beta$ leads to an improved version of the $P_E$ criterion, offering a physically grounded alternative. Using analytical derivations and Prairie Meteor Network data, the authors demonstrated that $y_{ter}=\ln(2\alpha\beta)$ preserves continuity in the allowed values for ablation coefficient and bulk density and provides a more comprehensive formulation that does not necessitate any initial assumptions of the not well-understood luminous efficiency, as the photometric mass in Eq. \ref{eq:PE} requires. This formulation is advantageous as it enhances the robustness and versatility of fireball classification methods, facilitating more accurate and efficient analysis of fireball phenomena without the need for extensive prior information, such as empirical datasets needed to define preset ablation coefficient values for strictly imposed fireball categories as well as no necessity to define sharp cut-off boundary values between the groups. Eliminating the requirement for predefined parameters, such as bulk densities and ablation coefficients, which were made default within the $P_E$ groups (Table \ref{tab:fireball_classification}), allows for a more physical, continuous, and adaptable approach to fireball classification, ensuring a more adequate understanding of meteoroid properties and behaviors. This underscores, once more, the advantage of reducing the number of unknowns as described at the beginning of this section.

By determining the dependence on mass and velocity, \citeA{Borovicka2022AA_II} defined a new parameter for the evaluation of material strength based on the maximal dynamic pressure for large meteoroids, the so-called pressure resistance factor or simply the pressure factor, abbreviated as $Pf$. It is defined as:

\begin{equation}
Pf = 100  \rho_{max}  \cos z^{\ -1}  M_{phot}^{\ -1/3}  V_{beg}^{\ -3/2},
\label{eq:Pf}
\end{equation}

where \( \rho_{max} \) is the maximal dynamic pressure in $MPa$. Five categories for $Pf$ classification, designated as $Pf$-I to $Pf$-V (from the strongest to the weakest), are established, with the respective boundary $Pf$ values delineated as follows: $Pf$ greater than 0.85 for $Pf$-I, between 0.27 and 0.85 for $Pf$-II, between 0.085 and 0.27 for $Pf$-III, between 0.027 and 0.085 for $Pf$-IV, and less than or equal to 0.027 for $Pf$-V.

Typically, fireballs associated with meteorite-dropping events are expected to be classified as $P_E$ type I or $Pf$-I. Alternatively, $\alpha$-$\beta$, in addition to the terminal mass computation explained earlier, also provide a visual classification defining regions that categorize the likelihood of meteorite fall: ``Likely fall'', ``Possible fall'', and ``Unlikely fall'' \cite{Sansom2019ApJ, moreno2020physically, Boaca2022ApJ}. A terminal mass of 0.05 kg is commonly recognized as a threshold for meteorite production \cite{halliday1996detailed}, which is a crucial factor in determining the potential for a meteoroid to survive atmospheric entry and reach the Earth's surface. In practice, this threshold may be adjusted to accommodate the realistic chances of locating a meteorite of similar size within a given terrain. The ``Possible Fall'' region is defined by an upper boundary curve obtained analytically and corresponding to scenarios without meteoroid rotation, denoted as $\mu=0$, and a lower boundary is specified by the maximum value of $\mu=2/3$ \cite{Sansom2019ApJ}. 

The applicability of using $\alpha$ and $\beta$ to classify meteoroid and asteroid impacts at large was initially introduced by \citeA{Gritsevichcite2009Classification, gritsevich2011DokPh}. Depending on these values, potential outcomes include the formation of a single massive crater, atmospheric destruction with multiple resulting fragments forming craters on Earth, ablation with smaller fragments reaching Earth without producing craters, or complete destruction and evaporation in the atmosphere (e.g., Tunguska event), see \citeA{Gritsevich2012CosRe, Turchak2014JTAM}.

\section{Inferring atmospheric flight parameters} \label{sec:inferring}

The theory on atmospheric flight characterization described above underscores the need for velocity measurements as a function of height to elucidate the dynamics of meteoroid entry into the Earth's atmosphere. Our goal, however, is to infer the velocity profile and mass loss rates without explicitly using the complete observational data, i.e. when observations are incomplete or when such data are not available. Specifically, in this work we aim to achieve this using only two points (initial and terminal) of the atmospheric trajectory, with available values of height, velocity, and mass from the EN catalog. In the context of this paper, terminal velocity refers to the last measured velocity point, while terminal height is the height where the last measured velocity point is provided. Initial velocity, on the other hand, represents velocity at the provided starting point of the fireball, after accounting for atmospheric drag. This value, as defined in \citeA{Borovicka2022AA_I}, corresponds to the velocity the meteoroid would have at that position if it were influenced solely by gravity, excluding the effects of atmospheric forces.

\subsection{The EN fireball catalog}

As a study case, we employ the catalog comprising 824 fireballs, as observed and analyzed by the EN from 2017 to 2018 \cite{Borovicka2022AA_I, Borovicka2022AA_II}. Despite the comprehensive nature of this dataset, including a multitude of events and computed parameters, it lacks the raw observational data and other derived information such as the velocity profiles. Nevertheless, the dataset provides, among others, initial and terminal values of the characteristic parameters of the luminous phase, alongside integrated energy and classifications using $P_E$ and $Pf$. 

To convert pixel positions into azimuths and zenith distances, the all-sky conversion formulas developed by \citeA{Borovi1995} were utilized. These formulas employ a double exponential function to characterize radial lens distortion, though refraction was not explicitly accounted for. For IP cameras or other devices capturing only a portion of the sky, gnomonic conversion formulas from \citeA{Borovicka2014}, which incorporate polynomial lens distortion, were applied. The fireball's atmospheric trajectory was determined based on positional measurements from all cameras. \citeA{Borovicka1990} straight least squares method was used for this purpose, which presumes that the trajectory is a straight line in space. The trajectory was calculated through an iterative process that minimized the miss distances of individual lines of sight from the trajectory.

After computing the straight trajectory, shutter break measurements were employed to determine the fireball's position on the trajectory as a function of time. Each measurement was projected onto the trajectory, and the distance of the projected point from the trajectory's starting point was calculated, referred to as the length. Thus, the length as a function of time was obtained. Two fits were conducted during the analysis procedure. The first fit encompassed the entire fireball, correlating time and height, which is crucial for integrating light curves observed as a function of time. This fit also determined the velocity at maximum brightness and the end of the fireball's visibility. The ablation coefficient was fitted as a free parameter if the data quality was high. Conversely, if the data quality was less satisfactory, the ablation coefficient was set to a fixed value (typically 0.02 $s^2\,km^{-2}$), and the other parameters were fitted.

However, fitting the entire fireball trajectory this way may not effectively facilitate backward extrapolation to determine the pre-atmospheric velocity, which is essential for calculating the orbit \cite{dmitriev2015orbit}. This inadequacy could arise because the fit does not account for fragmentation, leading to underestimation of the dynamic mass and overestimating the initial velocity. To address this, \citeA{Borovicka2022AA_I} conducted a second fit using only the fireball trajectory's initial segment, before the fragmentation's onset. The ablation coefficient was fixed due to the negligible deceleration at this early stage. The mass was also fixed, based on values obtained from photometry by integrating the radiated energy over the fireball's trajectory, under the assumption that the radiated energy is proportional to the kinetic energy lost due to ablation. It is worth noting that in the cataloged data $M_{beg}=M_{pho}$, while the $M_{ter}$ was computed dynamically under the assumption of $\rho_m = 3000 \,\,kg\,m^{-3}$ and $(1/2)c_d A=0.7$ \cite{Borovicka2022AA_I}.

\subsection{Optimization problem}

To infer fireball parameters, we use the Differential Evolution (DE) algorithm for global optimization, a metaheuristic method built on the foundational work of \citeA{Storn1997JGOpt}. DE is a population-based stochastic optimization technique that iteratively improves candidate solutions regarding a given objective function. DE operates through a simple cycle of mutation, crossover, and selection processes applied to a population of candidate solutions. We implement a modification for improved convergence speed, which allows trial vectors to immediately capitalize on improved solutions \cite{Wormington1999RSPTA}. At each iteration, the algorithm modifies each potential solution by integrating it with other potential solutions within the population to form a trial candidate \cite{Qiang2014}. This process can be formally described as follows.

Let $x_i^{(t)}$ denote the $i$-th candidate solution in the population at iteration $t$. The DE algorithm generates a trial vector $u_i^{(t+1)}$ by selecting three distinct vectors $x_{r1}^{(t)}$, $x_{r2}^{(t)}$, and $x_{r3}^{(t)}$ at random from the current population. The trial vector is computed using the following equation:

\begin{equation}
u_i^{(t+1)} = x_{r1}^{(t)} + F \cdot (x_{r2}^{(t)} - x_{r3}^{(t)}),
\end{equation}

where $F$ is a scaling factor that controls the amplification of the differential variation $(x_{r2}^{(t)} - x_{r3}^{(t)})$. 

The effectiveness of DE stems from its ability to balance exploration and exploitation of the search space. The mutation strategy introduces variation and examines new areas, while crossover and selection refine solutions toward the optimum. It is particularly well-suited for optimization problems where the landscape of the objective function is complex. For further details on the implementation and parameters of the differential evolution algorithm in computational practice, refer to the documentation of the \textit{SciPy} library \cite{2020SciPy-NMeth}.

Our optimization problem involves using two points from the luminous trajectory phase — the initial and terminal points ($V$, $h$, and $M$ available from the catalog) — as input to determine the optimal values of output parameters $\alpha$, $\sigma$, and $\rho_m$. These output parameters enable us to accurately reconstruct the velocity profile based on a purely dynamic model. 

As discussed in \citeA{Gritsevich2007SoSyR, Gritsevich2008Parameters}, a weighted least squares method can be applied to analyze the height and velocity data observed, utilizing Eq. \ref{eq_y} to determine the $\alpha$ and $\beta$ parameters. \citeA{Gritsevich2008Parameters} proposes that adopting an exponential form of Eq. \ref{eq_y} enhances computational performance. This improvement is attributed to the singularity of Eq. \ref{eq_y} at $V=V_{beg}$ and the observation that the height and velocity of a meteoroid decrease as it approaches the Earth's surface (hence the function is otherwise well-defined and monotonous when $V \neq V_{\text{beg}}$). In mathematical terms, this approach shifts from a direct comparison of observed versus predicted by theory height values, represented by simply $y$, to a comparison between $e^{-y}$ values. However, in this study, we choose to solve the initial height numerically from Eq. \ref{eq_y}, just to ease the residuals operations. In each iteration, $\beta$ is directly computed by Eq. \ref{eq_beta_abla}, and the initial and terminal masses are estimated using Eq. \ref{eq:mass_init} and Eq. \ref{eq:mass_terminal}, respectively. With all this, we define the objective function to minimize as:


\begin{equation}
\begin{aligned}
    Q : \mathbb{R}^3 &\rightarrow \mathbb{R}\\
    Q(\alpha, \sigma, \rho_m) &= \mathfrak{r}(h_{beg}, V_{beg}, M_{beg}, h_{ter}, V_{ter}, M_{ter}),
\end{aligned}
\label{eq:my_equation}
\end{equation}

where $\mathfrak{r}^{(t)}$ is the sum of absolute residuals in each iteration:



\begin{equation}
    \mathfrak{r}_{v_{beg}}^{(t)} =  y^{-1}_{beg}(v_{beg};\alpha^{(t)},\beta^{(t)}) - 1, \label{eq_r_v_beg}
\end{equation}



\begin{equation}
    \mathfrak{r}_{y_{ter}}^{(t)}=\ln 2\alpha^{(t)} +\beta^{(t)} -\ln \left (\overline{E}i(\beta^{(t)} )-\overline{E}i\left(\beta^{(t)} \left(\frac{V_{ter}}{V_{beg}}\right)^2\right)\right) - \frac{h_{ter}}{h_{0}},   \label{eq_r_y_ter}
\end{equation}


\begin{equation}
   \mathfrak{r}_{M_{beg}}^{(t)} = \left(\frac{1}{2} \frac{ \rho_{sl} h_0}{\alpha^{(t)} \left(\rho_m^{(t)}\right)^{2/3} \sin\gamma}\right)^{3} - M_{beg},\label{eq_r_m_beg}
\end{equation}

\begin{equation}
    \mathfrak{r}_{M_{ter}}^{(t)} = M_{beg}^{(t)} e^{-\frac{\beta^{(t)}}{1-\mu} \left(1-\left(\frac{V_{ter}}{V_{beg}}\right)^2\right)}  - M_{ter},\label{eq_m_y_ter}
\end{equation}

to finally compose:

\begin{equation}
    \mathfrak{r}^{(t)} = c_1\left| \mathfrak{r}_{v_{beg}}^{(t)} \right| + c_2\left| \mathfrak{r}_{y_{ter}}^{(t)} \right| + c_3\left| \mathfrak{r}_{M_{beg}}^{(t)} \right| + c_4\left| \mathfrak{r}_{M_{ter}}^{(t)} \right|,  \label{eq_r}
\end{equation}

where constants $c_{i}$ are used to equally weight the residuals according to the desired residual conditions to be achieved. Note that the variables $v$ and $y$ are mutually dependent, leading to implicit equations. The initial velocity is solved numerically, whereas the initial height is not explicitly resolved. Conversely, the terminal height is determined analytically, while the terminal velocities are not directly computed. These residuals are used in the optimization process; however, the residuals for the initial height and terminal velocity will also be evaluated subsequently for general analysis.
 
The following physical constraints are applied during the minimization process:

\begin{equation}
\begin{aligned}
& \underset{(\alpha, \sigma, \rho_m) \in \mathbb{R}^3}{\text{minimize}}
& & Q(\alpha, \sigma, \rho_m), \\
& \text{subject to}
& & 
\begin{cases}
\alpha > 0,\\
\sigma_l \leq \sigma \leq \sigma_u,\\
\rho_{m,l} \leq \rho_m \leq \rho_{m,u}.
\end{cases}
\end{aligned}
\end{equation}

We adopt $c_d A = 1.8$, a commonly used value describing realistic complex-shaped objects \cite{Gritsevich2008SoSyR, Gritsevich2024}. We also set $\mu = 2/3$, as it represents the most likely shape change coefficient for meteoroid atmospheric entry scenarios \cite{Bouquet2014PSS}. Unfortunately, there is no inverse function for the exponential integral, which can be analytically expressed. This necessitates the adoption of numerical methods to approximate the solution. Therefore, to invert Eq. \ref{eq_y} and estimate the initial velocity in each iteration (Eq. \ref{eq_r_v_beg}), we utilize Nelder-Mead's method \cite{Gao2012}. 

In the optimization process, we have established a maximum iteration limit of 1000 for each optimization run. To robustly estimate the variation in the optimization outcomes and assess the stability of our approach, each optimization case is repeated 100 times. We select the best solution from these repetitions as the one with the lowest overall residuals. We consider the fit as not converged if the terminal height deviates by more than 100 $m$, the velocities by more of 100 $m\,s^{-1}$, or the masses by more than 1 $g$.

In our research, we conduct two distinct calculations. First, a set of calculations specifically restricted to the meteoroid vulk density and ablation coefficient with a 5\% of margin, based on the $P_E$ criterion. Ablation coefficient and meteoroid bulk density are summarized in Table \ref{tab:fireball_classification} according to each type. Iron meteoroids are characterized by a density range of 6000 to 8000 $kg\,m^{-3}$ and an ablation coefficient ranging from 0.01 to 0.07 $s^2\,km^{-2}$ \cite{Ceplecha1998SSRv, Vida2018MNRAS, Vojavcek2020PSS}. The second set of calculations is not constrained by the $P_E$ classification. The ablation coefficient is allowed to vary widely from 0.005 $s^2\,km^{-2}$ to 0.25 $s^2\,km^{-2}$, and the meteoroid density from 270 $kg\,m^{-3}$ to 8000 $kg\,m^{-3}$, with the same 5\% of margin for the limits.

\begin{table}[H]
\centering
\caption{Meteoroid bulk density and ablation coefficient for different types of fireballs suggested based on the $P_E$ criterion \cite{Ceplecha1998SSRv}.}
\label{tab:fireball_classification}
\begin{tabular}{lcc}
\hline
Type & $\sigma$\ [$s^2\,km^{-2}$] &  $\rho_m$ [$kg\,m^{-3}$] \\
\hline
I & 0.014 & 3700 \\
I/II & 0.014 -- 0.042 & 3700 -- 2000 \\
II & 0.042 & 2000 \\
IIIA & 0.1 & 750 \\
IIIA/B & 0.1 -- 0.12 & 750 -- 270 \\
IIIB & 0.21 & 270 \\
\hline
\end{tabular}
\end{table}

Some of the iron meteoroids were initially flagged through spectral measurements and are thus included in the database. Others, suspected to be iron due to their atmospheric flight characteristics and light curves but lacking available spectra, are detailed in Table 1 of \citeA{Borovicka2022AA_II}.

\subsection{Performance of the fitted parameters}

To evaluate the fitted parameters, in addition to comparing some of our reconstructed velocity profiles with point-by-point measured velocities for five events (the only ones available to us), we compare the performance of our $\alpha-\beta$ results in predicting terminal height against databases in which these parameters were derived using all measured points along the atmospheric trajectory. Particularly, we utilize the database compiled by \citeA{Gritsevich2009AdSpR}, based on observations of 143 Meteorite Observation and Recovery Project (MORP) fireballs \cite{halliday1996detailed}. As the residuals from the fit were not disclosed, we rely on the predictive performance of certain analytical approximations \cite{Gritsevich2016} to estimate the terminal height as a function of $\alpha$, $\beta$, and $V_{ter}$ \cite{Moreno2015Icar}. These formulas are as follows:

\begin{equation}
h_I=h_0 \cdot y_{ter}^{I}=h_0 \cdot \ln (2 \alpha \beta), \label{eq_hI}
\end{equation}

\begin{equation}
h_{II}=h_0 \cdot y_{ter}^{II}=h_0 \cdot \ln \left(\frac{2 \alpha \beta}{\left(1-e^{\beta\left(\left(\frac{V_{ter}}{V_{beg}}\right)^2-1\right)}\right)}\right), \label{eq_hII}
\end{equation}

\begin{equation}
h_{III}=h_0 \cdot y_{ter}^{III}=h_0 \cdot \ln \left(\frac{2 \alpha(\beta-1.1)}{\left(1-e^{(\beta-1.1)\left(\left(\frac{V_{ter}}{V_{beg}}\right)^2-1\right)}\right)}\right) \label{eq_hIII}.
\end{equation}

Using dimensionless formulas for terminal height $y_{ter}$, one can predict the dimensional terminal height $h_{ter}$ under various assumptions. By neglecting deceleration, the terminal height can be calculated using Eq. \ref{eq_hI}, derived as an approximation to the analytical solution valid for $\beta \gg 1$ \cite{Gritsevich2008DokPh}. Note that this simplification does not guarantee accurate predictions for not decelerating fireballs, as other mechanisms, such as complex fragmentation, may influence the outcome. Other solution that accounts for some deceleration is provided by Eq. \ref{eq_hII}. Finally, Eq. \ref{eq_hIII} presents an approximation that introduces a shift as a function of $\beta$ and $v$ and provides a closer match to the analytical solution \cite{Gritsevich2016}.

\section{Results and discussion} \label{sec:results}

\begin{figure}[!t]\centering
  \includegraphics[,width=1\linewidth]{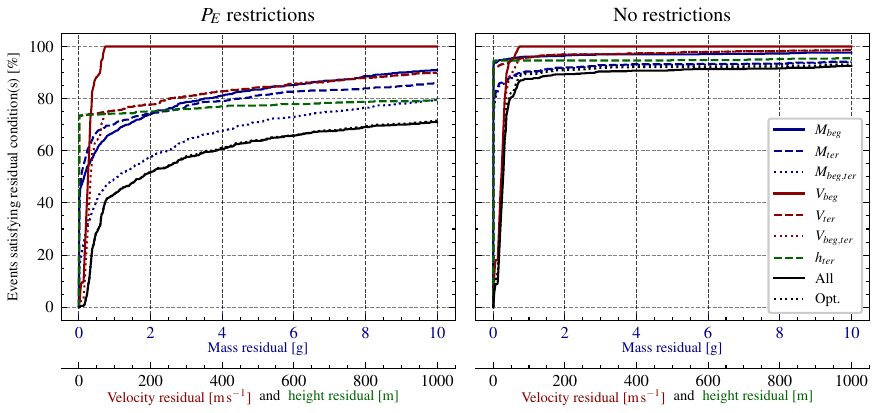}
  \caption{Percentage of fitted events satisfying the residual conditions as a function of the residual thresholds, with $P_E$ restrictions (left) and without restrictions (right). The y-axis represents the percentage of events meeting the residual conditions, while the x-axis shows the residual thresholds. Velocity and height are represented on the same axis but with different units. The final label in the legend corresponds to the residual conditions applied during the optimization process.}

  \label{fig:wrongs_dependency}
\end{figure}

Figure \ref{fig:wrongs_dependency} presents a comparison of the percentage of fitted events that satisfy the residual conditions as a function of varying residual thresholds, for scenarios both with and without the $P_E$ fireball type constraint, based on the analysis of 824 EN fireballs. The contribution of each residual can be analyzed separately, with the initial mass being the most significant in both cases. In contrast, the initial velocity appears to be the most easily satisfied parameter. The unrestricted fits rapidly stabilize in high percentages of events meeting the conditions, whereas the fits with $P_E$ restrictions exhibit a much slower convergence. For fits with $P_E$ restrictions, even when allowing residuals up to 1 $km\,s^{-1}$ in velocity, 1 $km$ in terminal height, and 10 $g$ in mass, around 70\% of events satisfy these conditions. In the case of the fits without restrictions, for residuals with lower masses of 1 $g$, velocities of 100 $m\,s^{-1}$, and terminal heights of 100 $m$, 90\% of fitted events meet these criteria, in comparison with 44\% of the restricted fits. Fig. \ref{fig:wrongs_dependency} also demonstrates that explicitly including terminal velocity in the optimization has minimal impact on the residuals. This parameter was omitted in favor of terminal height, as the two are inherently coupled, as it happens with the initial height.


\begin{figure}[!t]\centering
  \includegraphics[width=.8\linewidth]{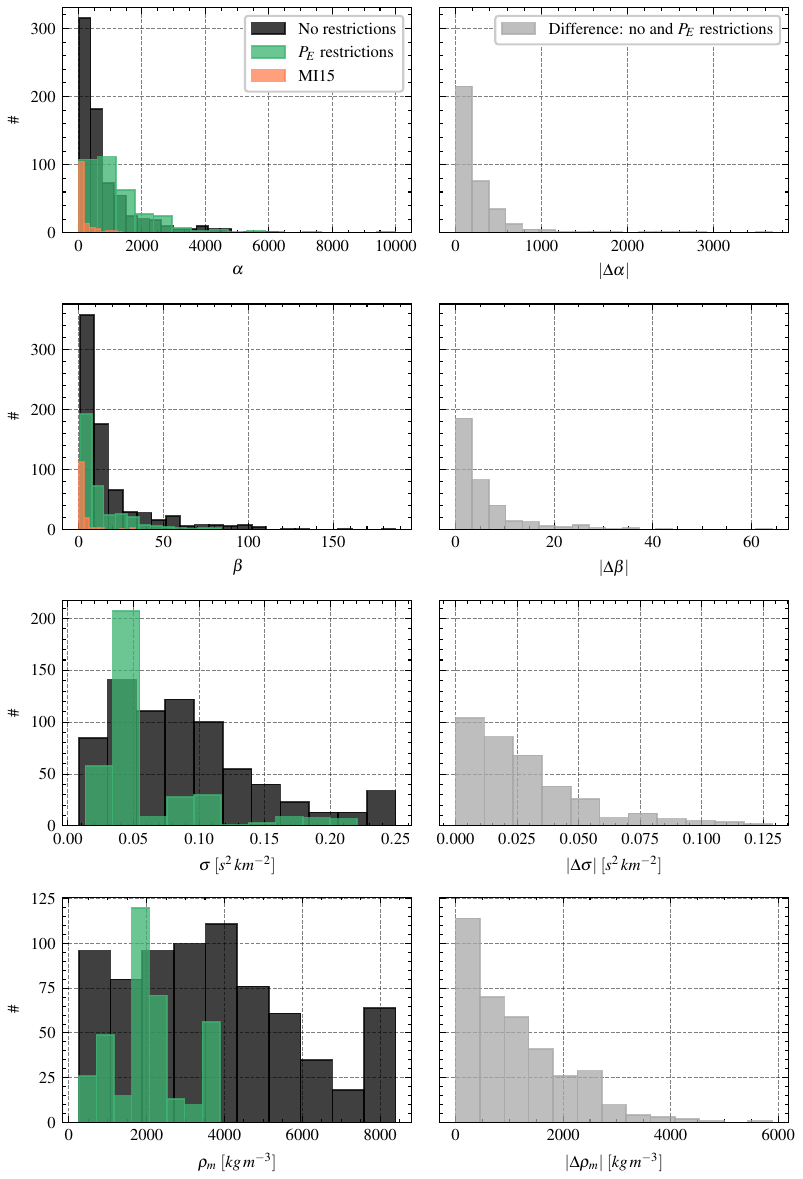}
  \caption{Histograms display the fitted parameters $\alpha$, $\sigma$, and $\rho$, with $\beta$ being implicitly derived from these for the EN catalog. Fit results that meet residual conditions are presented with and without constraints based on the fireball type following the $P_E$ criterion. The second column illustrates the absolute difference in outcomes. MI15 denotes the values reported by \citeA{Moreno2015Icar} for comparison, although these are provided only for $\alpha$ and $\beta$ derived from the MORP database, which were conventionally obtained using the velocity profiles described by \citeA{Gritsevich2009AdSpR}. Histogram bins are computed with Scott's rule \cite{Scott1979}.}
  \label{fig:parameters_fit_diff}
\end{figure}

Fig. \ref{fig:parameters_fit_diff} presents a comparative histogram analysis of fitted parameters for the two cases: one without any constraints and the other under constraints imposed by the fireball type as per the $P_E$ criterion. The distributions of both $\alpha$ and $\beta$ for the two cases are predominantly concentrated at values of $\alpha$ less than 1000 and $\beta$ less than 50. Additionally, Fig. \ref{fig:parameters_fit_diff} shows $\alpha$ and $\beta$ values estimated using all observed points from the MORP database \cite{Moreno2015Icar}, although no mass constraints were applied. The $\alpha$ and $\beta$ histograms reveal moderate difference, while the meteoroid bulk density histogram shows drastic difference between the two fits. The unconstrained scenario shows a peak of $\sigma$ values around 0.03--0.04 $s^2\,km^{-2}$ and a concentration of $\rho_m$ values consistent with asteroid-like compositions, with even higher strength values observed for a substantial number of events. Under $P_E$ restrictions, the most frequent values of $\alpha$ are not the smallest, in contrast to the unconstrained fits.


\begin{figure}[!t]\centering
  \includegraphics[width=0.75\linewidth]{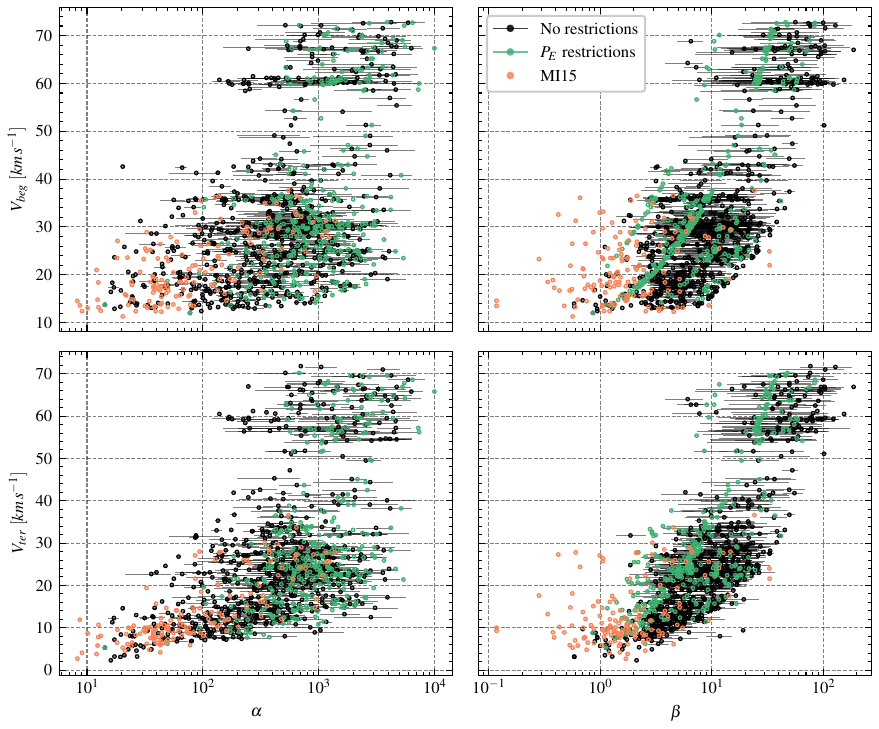}
  \caption{Fitted values of $\alpha$ and $\beta$, both with and without the $P_E$ type constraint, plotted against initial and terminal velocities. Only events meeting the criteria of a terminal height deviation within 100 $m$ or a masses deviations within 1 $g$ are included. The converged solutions standard deviations are depicted as solid gray straight lines. MI15, corresponds to \citeA{Moreno2015Icar} values, represents the MORP $\alpha$ and $\beta$ fitted conventionally with the velocity profiles by \citeA{Gritsevich2009AdSpR}.}
  \label{fig:MI15comp}
\end{figure}

To further compare our results with the parameters calculated for the MORP database using only (but the entire) velocity profiles \cite{Gritsevich2009AdSpR, Moreno2015Icar}, we present in Fig. \ref{fig:MI15comp} our fitted values of $\alpha$ and $\beta$ as a function of \(V_{beg}\) and \(V_{ter}\), both with and without the $P_E$ constraint. This figure only includes the events meeting residual conditions. The MI15 markers represent the MORP values, as fitted by \citeA{Moreno2015Icar} following the velocity profile interpretation approach of \citeA{Gritsevich2009AdSpR}. $\alpha$ values are slightly higher when no constraints are applied, while $\beta$ values decrease compared to the unrestricted fit. In MI15, the sample appears biased towards slower likely asteroidal impactors (primarily $Pf$-I and $Pf$-II). Although there is some agreement in $\alpha$ values, $\beta$ values show more significant differences. Our $\beta$ values are generally higher, and a data structure becomes evident when plotted against velocity, which also occurs to some extent with $\alpha$. This $\beta$ structure does not appear in MI15, and may result from the masses constraints applied in this work during the fitting process, which were absent in \citeA{Moreno2015Icar}. However, a direct comparison is limited to $Pf$-I and $Pf$-II events, as the published MORP database is biased toward slower impactors.

\begin{figure}[!t]\centering
  \includegraphics[width=\linewidth]{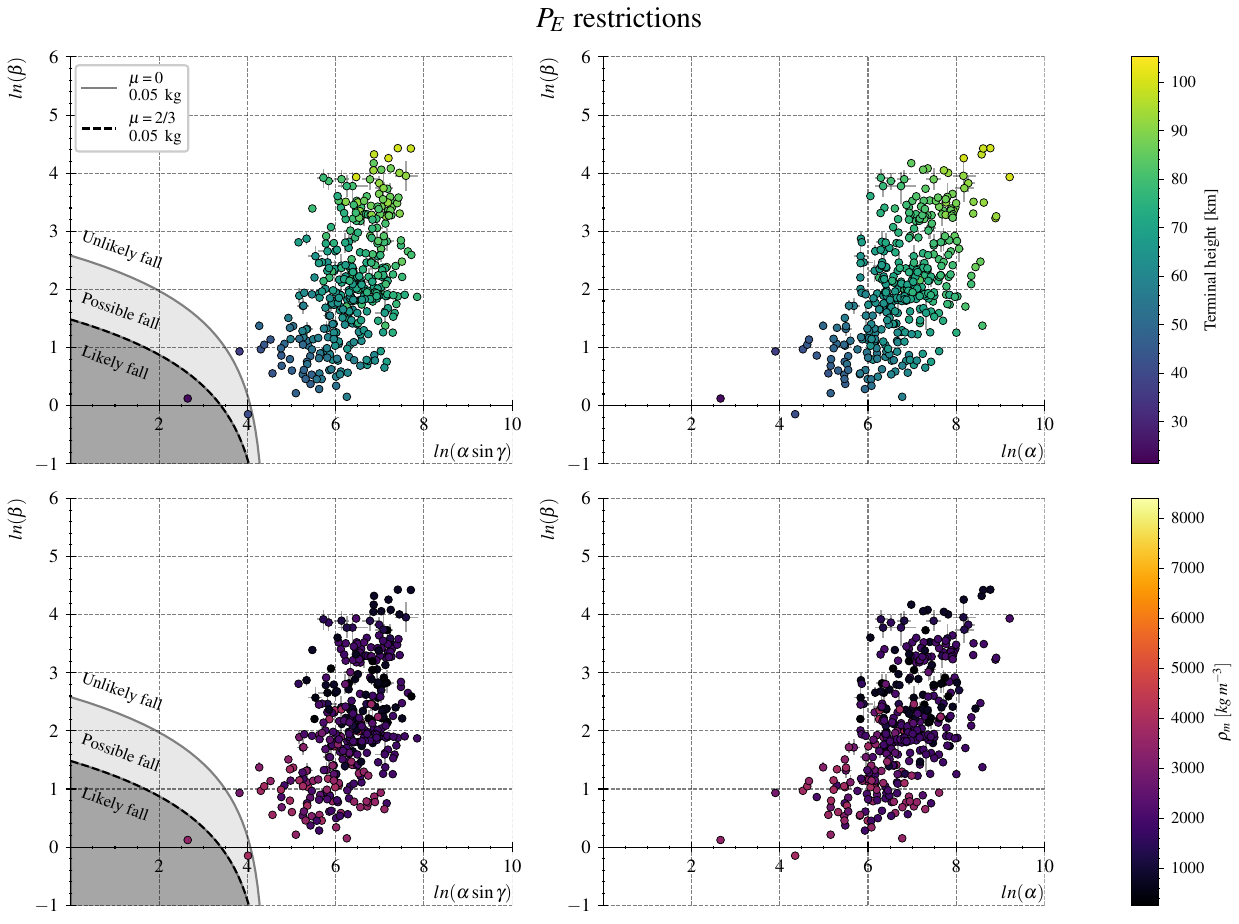}
  \caption{Distribution of fitted $\alpha$-$\beta$ that meet residual conditions, colored by terminal height (top row) and meteoroid bulk density (bottom row) for the EN catalog. Left column shows the distribution with trajectory slope dependence removed. Restrictions were applied based on the $P_E$ criterion. The converged solutions standard deviations are depicted as solid gray straight lines. The threshold curves for meteorite production correspond to a terminal mass of 0.05 kg for objects with a bulk density of 3500 $kg\,m^{-3}$.}
  \label{fig:abg_restrictions}
\end{figure}

\begin{figure}[!t]\centering
  \includegraphics[width=\linewidth]{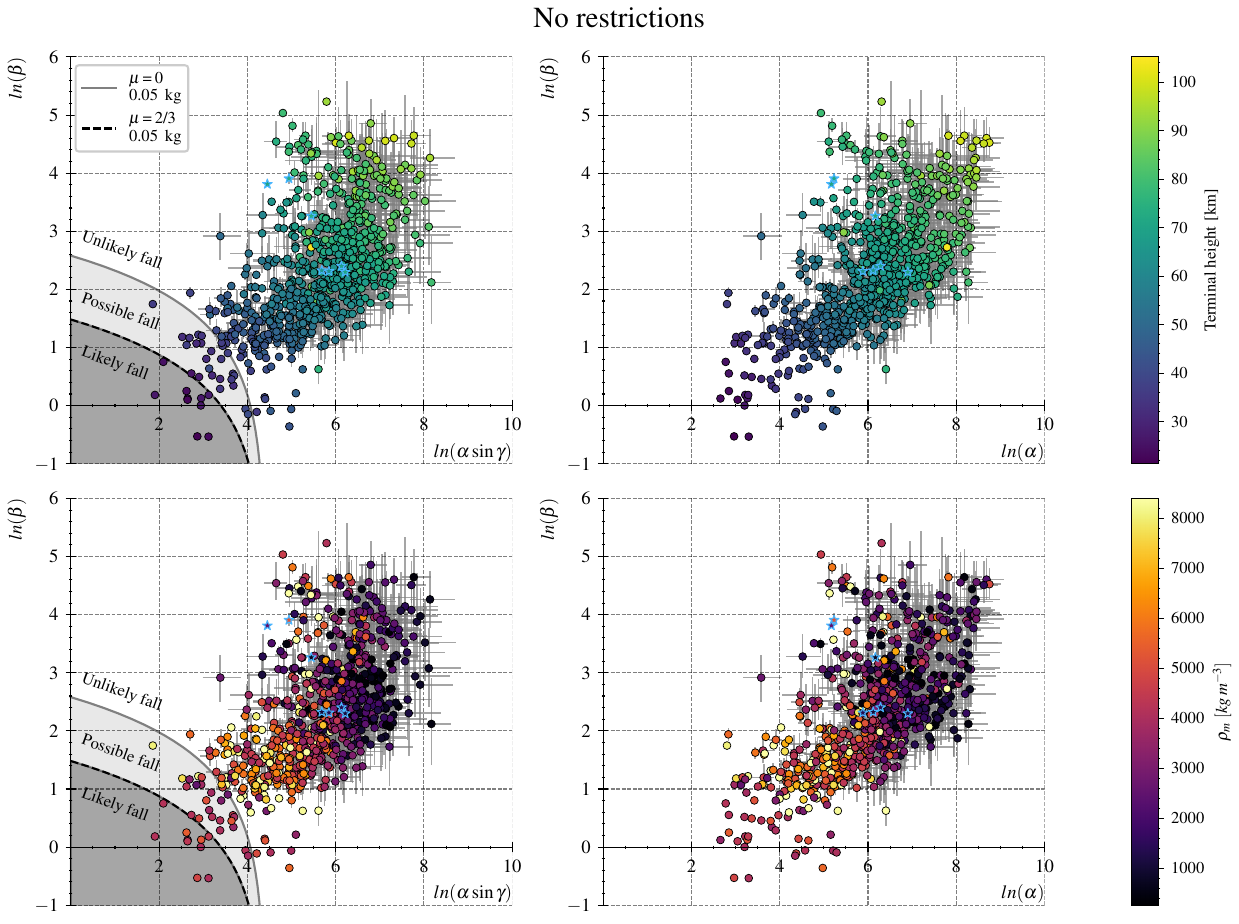}
  \caption{Distribution of fitted $\alpha$-$\beta$ that meet residual conditions, colored by terminal height (top row) and meteoroid bulk density (bottom row) for the EN catalog. Left column shows the distribution with trajectory slope dependence removed. No restrictions were used based on the $P_E$ criterion. The converged solutions standard deviations are depicted as solid gray straight lines. The threshold curves for meteorite production correspond to a terminal mass of 0.05 kg for objects with a bulk density of 3500 $kg\,m^{-3}$. Earlier suspected iron meteoroids are marked with a cyan star \cite{Vojavcek2020PSS, Borovicka2022AA_II}.}
  \label{fig:abg_no_restrictions}
\end{figure}

Fig. \ref{fig:abg_restrictions} illustrates the distribution of meteoroid parameters $\alpha$ and $\beta$ for the $P_E$ restricted fits, with the natural logarithms of these parameters (right) and after removing the dependency on the trajectory slope (left). The data points are color-coded to represent the terminal heights and the meteoroid bulk densities. Similarly, Fig. \ref{fig:abg_no_restrictions} shows the meteoroid parameters $\alpha$ and $\beta$ for fits without restrictions. Earlier suspected iron meteoroids \cite{Vojavcek2020PSS, Borovicka2022AA_II} are marked with a cyan star. These representations enable a visual classification of events based on their likelihood of yielding a recoverable meteorite. For both fit scenarios, only events that meet residual conditions are shown. As anticipated, events likely to result in meteorite production are characterized by lower terminal heights and densities consistent with asteroidal objects. In Fig. \ref{fig:abg_restrictions}, one event (EN061118\_190859) stands out as an outlier in terminal height, significantly separated from other events with similar terminal heights. This event is a fast grazing event (\(\gamma = 5.51^\circ\)) with an initial velocity of \(V_{beg} = 66.476 \,\, km\,s^{-1}\), no observable deceleration, and a terminal height of 104.4 $km$. In the plots of the second row, a concentration of events with high estimated densities is evident. These events fall outside the region associated with potential meteorite production and are clearly separated from the population of weaker events.

The anticipated unclustering in Fig. \ref{fig:abg_restrictions} and \ref{fig:abg_no_restrictions} upon the removal of slope dependency did not materialize as previously seen in other databases. For example, in \citeA{Sansom2019ApJ, moreno2020physically}, plotting $ln(\alpha)$ versus $ln(\beta)$ revealed that the observed terminal height of a fireball typically decreases with a reduction in either or both $\alpha$ and $\beta$. However, these authors noted that this trend is influenced by the trajectory slope, which varies per event. When accounting for this by adjusting the $\alpha$-$\beta$ diagram with the term $\alpha \sin \gamma$ on the x-axis (where $\gamma$ represents the trajectory slope), the visual relationship with end heights dissipates. In our work, however, we do not observe this pattern, primarily because we calibrate $\alpha$ and $\beta$ to match the photometric mass, pointing to a possible inadequate estimation of the masses, a discrepancy between the photometric and dynamic approaches, or again to the fitting methods used.

\begin{figure}[!t]\centering
  \includegraphics[width=0.85\linewidth]{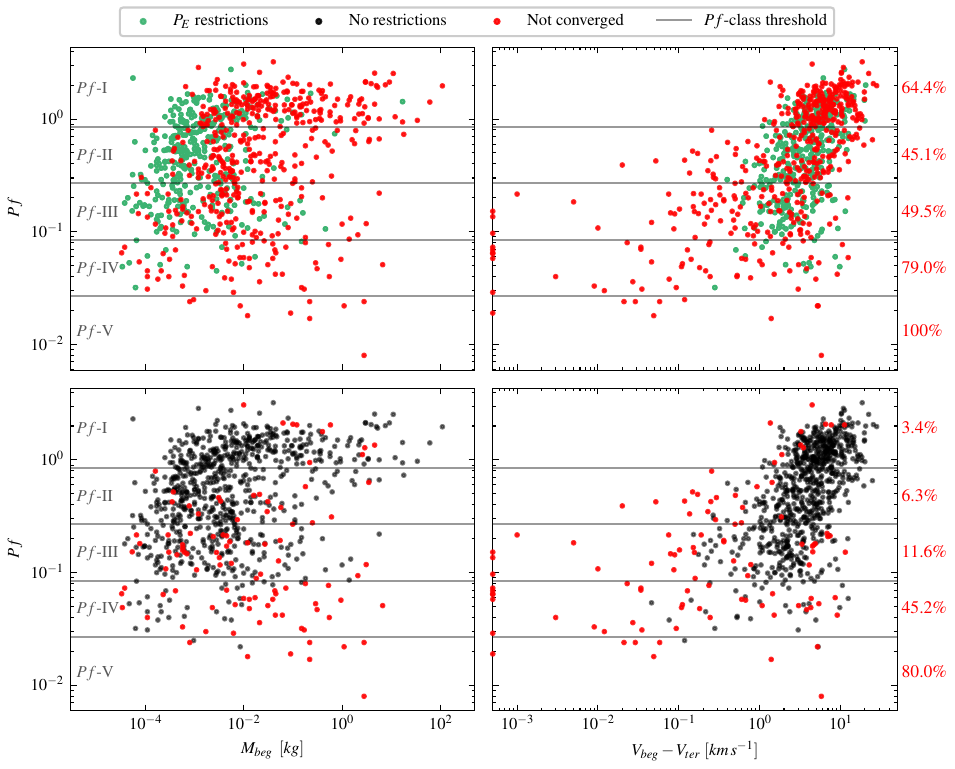}
  \caption{Scatter plots showing the classification of events based on the $Pf$-class against initial mass (\(M_{beg}\), left panels) and total deceleration (\(V_{beg} - V_{ter}\), right panels). The plots compare results for cases with $P_E$ restrictions (green points) and without restrictions (black points), and also denote the non-converged solutions (red points). Horizontal lines indicate $Pf$-class thresholds. On the right side of the panels, the percentage of events that do not satisfy the residual conditions is displayed for each $Pf$-class. Dots on top of the y-axis represent events without measurable deceleration.}
  \label{fig:wrongs}
\end{figure}

In our investigation of fireball events where it has not been possible to adequately reconstruct the atmospheric flight (when our residual conditions are not satisfied), we identify a certain correlation between some parameters. Figure \ref{fig:wrongs} presents scatter plots that classify events according to their $Pf$-class as a function of initial mass (\(M_{beg}\), left panels) and total deceleration (\(V_{beg} - V_{ter}\), right panels). The plots illustrate results for cases with $P_E$ restrictions (green points), without restrictions (black points), and non-converged solutions (red points). Horizontal lines mark the $Pf$-class thresholds, while the percentages on the right side indicate the proportion of events that fail to satisfy the residual conditions. The data reveal a clear trend: the lower the $Pf$, indicating a greater propensity to fragment, the higher the percentage of fits that do not meet the residual requirements. This trend is particularly noticeable in the unconstrained case, as shown by the percentages in red on the right-hand side of the plots. For the restricted case, it can be observed that the non-converged fits are more concentrated at larger masses, a tendency that persists but is less pronounced in the unrestricted case. Regarding total deceleration, events with smaller (or negligible) velocity differences between initial and terminal values show a higher failure rate in satisfying the residual conditions, indicating that the method struggles to converge under these circumstances.

\begin{figure}[!t]\centering
  \includegraphics[width=0.75\linewidth]{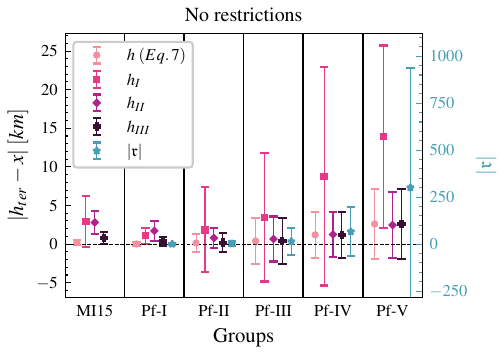}
  \caption{Mean values and standard deviations of residuals for $h=h_0y$ (Eq. \ref{eq_y}) and for terminal height formulas $h_I$, $h_{II}$, and $h_{III}$ (Eqs. \ref{eq_hI}, \ref{eq_hII}, and \ref{eq_hIII}, respectively). The right y-axis, highlighted in blue, depicts the residuals of the fit excluding the constraint based on fireball type according to the $P_E$ criterion. The data are categorized based on their classification via the $Pf$ parameter in the EN catalog. The first group, MI15, corresponds to \citeA{Moreno2015Icar} values, where the $\alpha$ and $\beta$ fit was conducted conventionally with the velocity profiles of the MORP database by \citeA{Gritsevich2009AdSpR}. $x$ denotes $h$, $h_I$, $h_{II}$, or $h_{III}$.
}
  \label{fig:pf_vs_res}
\end{figure}

Fig. \ref{fig:pf_vs_res} presents a comparative analysis of residual values for various terminal height formulas, specifically derived from Eq. \ref{eq_y}, and $h_I$, $h_{II}$, and $h_{III}$ as defined by their respective Eq. \ref{eq_hI}, \ref{eq_hII}, and \ref{eq_hIII}. The left y-axis shows the mean values and non-Gaussian standard deviations of residuals for these formulas, while the right y-axis, colored in blue, represents the residuals of the fit when the constraint based on fireball type, according to the $P_E$ criterion, is excluded from consideration. The data points are grouped into categories based on the $Pf$ parameter with five distinct groups. The group MI15 refers to the values obtained from \citeA{Moreno2015Icar}, where the $\alpha$ and $\beta$ parameters were fitted using the conventional velocity profiles provided by \citeA{Gritsevich2009AdSpR}. Remarkably, the events classified under the $Pf$-I group exhibit superior performance in predicting terminal height compared to the reference MI15 group, despite the latter having access to the complete point-by-point velocity profile. However, this is to some extent expected, as their fits attempted to satisfy all data points, while our approach concentrates on just two with explicit focus on minimizing the residuals in terminal height under our defined residual conditions. The predictive accuracy diminishes towards the $Pf$-V group. This group is characterized by fast impact, likely involving ice agglomerates, and the most fragile impactors, indicative of complex fragmentation patterns or low observed deceleration. Such conditions may present a challenge for the developed dynamic model, leading to increased deviations in the predicted terminal heights. The performance in predicting the terminal height of $h_I$ noticeably deteriorates more than other parameters as the $Pf$ value decreases.

\begin{figure}[!t]\centering
  \includegraphics[,width=0.5\linewidth]{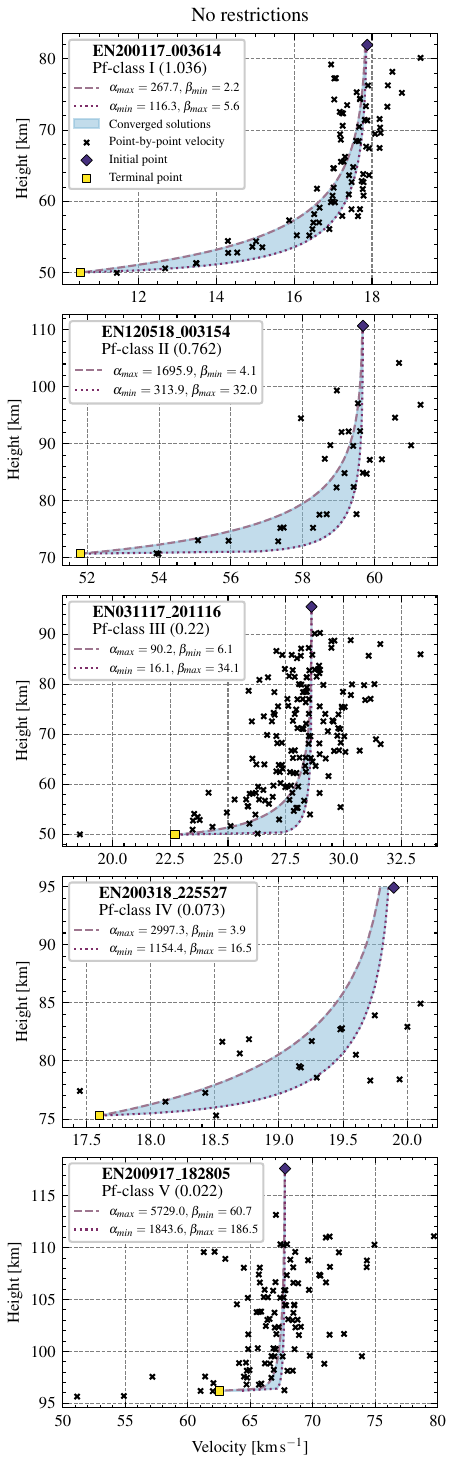}
  \caption{Reconstructed velocity profile for five events with different $Pf$-classes. The $\alpha$-$\beta$ converged solution parameter space is shown. Point-by-point velocity values were provided in personal communication by Jiří Borovička.}
  \label{fig:h_vs_vel}
\end{figure}

Figure \ref{fig:h_vs_vel} presents example of velocity profiles reconstructed for five distinct events, each classified into one of the different categories based on the $Pf$ criterion. These five examples are provided for demonstration purposes only, as they represent the complete set of cases available to us via personal communication with Jiří Borovička. The parameter space of the velocity profile for the solutions that satisfy the residual conditions is represented in light blue, bounded by the minimum and maximum $\alpha$-\(\beta\) combinations. The panels illustrate the best fits achieved through our optimization process, which was conducted without applying the $P_E$ criterion to maintain an unbiased assessment. As $Pf$ decreases, the alignment of the possible velocity profiles with observed points worsens. The velocity profiles are well constrained and align closely with the point-by-point velocity for the events of $Pf$-I and $Pf$-II class. Note that for the EN200318\_225527 fireball, only the velocity data for the second segment of its flight is available. The slight shift observed at the initial point may result from the cataloged initial velocity being treated as if no atmospheric deceleration had occurred, as meteoroids can decelerate by up to 500 $m\,s^{-1}$ prior to becoming detectable \cite{Vida2018MNRAS}. Although this assumption is generally valid for most cases given current instrument sensitivity, especially for meteoroids with the highest initial altitudes, it may not represent all scenarios.

A notable observation is the alteration in the classification of events when comparing the restricted and non-restricted fit models. Overall, 64\% of the fitted events with no restrictions that meet the residual conditions changed their $P_E$ type classification based on the estimated meteoroid bulk density and ablation coefficient, while 26\% are compatible with previous $P_E$ classification. This shift highlights the sensitivity of fireball categorization to the constraints applied during the fitting process and underscores the impact of model selection on interpreting atmospheric flight data. Adjusting fireball types due to model constraints may reflect underlying physical differences in the events or indicate a model-dependent bias.


Tables \ref{table:non_zero_terminal_mass_no_res} and \ref{table:non_zero_terminal_mass} present all events characterized by their non-zero terminal masses, which indicate their significance as meteorite-dropper candidates or potential impact hazards. The average computed values of $\alpha$, $\beta$, $\sigma$, and $\rho_m$ across all iterations during the optimization process that satisfied the residual conditions are reported. As expected, the parameters retrieved for events satisfying the residual conditions under $P_E$ constraints are virtually identical to those obtained without constraints, with all of them corresponding to $P_E$ types I/II. In some cases, the calculated densities appear to be unusually high (see Table \ref{tab:fireball_classification} for reference values). This can be attributed to the sensitivity of our model to the chosen $c_d A$, where lower values result in reduced bulk density estimates. These outcomes are also directly influenced by the masses provided in the catalog. It is worth noting that in addition to incorporating the initial mass estimates, variations in the dynamically calculated terminal masses in the catalog could also impact our results. As recognized in \citeA{Borovicka2022AA_I}, these terminal masses are approximate due to the assumptions made and the absence of a fragmentation model in their determination.


 \begin{table}
 \caption{Table of estimated parameters and residuals for events exhibiting non-zero terminal masses for the EN catalog. Restrictions were applied based on the $P_E$ criterion. The last four columns show the average parameters retrieved in this work. Only fits meeting residual conditions are listed.}
 \label{table:non_zero_terminal_mass_no_res}
 \centering
 \resizebox{\columnwidth}{!}{
 \begin{tabular}{lccccccccccccccc}
 \hline
Code & $h_{beg}$ & $V_{beg}$ & $M_{beg}$ & $h_{ter}$ & $V_{ter}$ & $M_{ter}$ & $P_E$ & $\alpha$ & $\beta$ & $\sigma$ & $\rho_m$  \\
 & $[km]$ & $[km\,s^{-1}]$ & $[kg]$ & $[km]$ & $[km\,s^{-1}]$ & $[kg]$ & [type] &  &   & $[s^2\,km^{-2}]$ & $[kg\,m^{-3}]$ \\
 \hline
EN130217\_210605 & 84.21 & 13.741 & 17 & 23.6 & 5.3 & 0.949 & I/II & 14.3 & 1.13 & 0.036 & 3466 \\
EN290518\_012837 & 72.69 & 12.034 & 0.23 & 41.0 & 8.8 & 0.069 & I/II & 77.5 & 0.86 & 0.036 & 3838 \\
EN170918\_185913 & 71.42 & 13.595 & 0.011 & 47.5 & 10.0 & 0.002 & I/II & 173.1 & 1.24 & 0.040 & 3453 \\
 \hline
 \end{tabular}
 }
 \end{table}

 \begin{table}
 \caption{Table of estimated parameters and residuals for events exhibiting non-zero terminal masses for the EN catalog. No restrictions were used based on the $P_E$ criterion. The last four columns show the average parameters retrieved in this work. Only fits meeting residual conditions are listed.} \label{table:non_zero_terminal_mass}
 \centering
 \resizebox{\columnwidth}{!}{
 \begin{tabular}{lccccccccccccccc}
 \hline
Code & $h_{beg}$ & $V_{beg}$ & $M_{beg}$ & $h_{ter}$ & $V_{ter}$ & $M_{ter}$ & $P_E$ & $\alpha$ & $\beta$ & $\sigma$ & $\rho_m$  \\
 & $[km]$ & $[km\,s^{-1}]$ & $[kg]$ & $[km]$ & $[km\,s^{-1}]$ & $[kg]$ & [type] &  &   & $[s^2\,km^{-2}]$ & $[kg\,m^{-3}]$ \\
 \hline
EN220117\_235255 & 86.46 & 23.166 & 0.46 & 35.2 & 9.4 & 0.006 & I & 53.1 & 1.73 & 0.019 & 4528 \\
EN260117\_201234 & 76.85 & 14.667 & 0.0065 & 43.5 & 9.7 & 0.002 & I & 144.8 & 0.7 & 0.019 & 5602 \\
EN020217\_014501 & 82.44 & 13.591 & 3.4 & 41.4 & 8.3 & 0.097 & I/II & 85.3 & 1.89 & 0.061 & 3737 \\
EN130217\_210605 & 84.21 & 13.741 & 17 & 23.6 & 5.3 & 0.949 & I/II & 14.3 & 1.13 & 0.036 & 3476 \\
EN230217\_194444 & 77.58 & 12.901 & 0.012 & 44.6 & 7.8 & 0.002 & I & 183 & 0.94 & 0.034 & 3873 \\
EN240217\_190640 & 84.22 & 17.816 & 2.2 & 31.9 & 5.4 & 0.005 & I & 30.7 & 2.23 & 0.042 & 5127 \\
EN270217\_023122 & 95.44 & 31.243 & 3.9 & 28.0 & 5.8 & 0.034 & I & 29 & 1.64 & 0.010 & 4921 \\
EN020317\_210949 & 82.55 & 17.686 & 0.045 & 38.6 & 9.7 & 0.004 & I & 83.6 & 1.15 & 0.022 & 4929 \\
EN250317\_233548 & 84.61 & 16.723 & 5.7 & 32.1 & 8.4 & 0.008 & I/II & 19.6 & 2.93 & 0.063 & 3779 \\
EN270317\_024041 & 80.90 & 15.983 & 0.57 & 37.4 & 7.7 & 0.012 & I & 65.6 & 1.68 & 0.039 & 3911 \\
EN290317\_195839 & 74.41 & 12.424 & 0.23 & 40.2 & 7.6 & 0.039 & I/II & 97.1 & 0.95 & 0.037 & 4181 \\
EN010417\_033446 & 82.03 & 13.864 & 0.067 & 45.4 & 9.3 & 0.001 & I & 110.8 & 2.57 & 0.080 & 5586 \\
EN160517\_205435 & 91.56 & 20.612 & 18 & 31.4 & 4.9 & 0.005 & II & 22.7 & 2.91 & 0.041 & 2501 \\
EN270517\_230125 & 87.70 & 20.256 & 0.71 & 32.8 & 7.7 & 0.019 & I & 45.7 & 1.41 & 0.021 & 3592 \\
EN130617\_211245 & 76.64 & 15.210 & 2.3 & 30.9 & 8.0 & 0.060 & I/II & 24.6 & 1.68 & 0.044 & 4107 \\
EN010717\_231011 & 86.19 & 25.934 & 1.8 & 37.3 & 9.6 & 0.003 & I & 54.7 & 2.47 & 0.022 & 2722 \\
EN220817\_211817 & 86.97 & 17.693 & 0.11 & 41.0 & 9.4 & 0.002 & I & 93.5 & 1.86 & 0.036 & 2958 \\
EN220817\_225359 & 81.18 & 13.710 & 0.045 & 45.8 & 8.8 & 0.009 & I & 197.9 & 0.91 & 0.029 & 4032 \\
EN081017\_183306 & 75.32 & 13.069 & 1.3 & 31.6 & 4.9 & 0.022 & I/II & 36 & 1.58 & 0.056 & 4169 \\
EN141117\_164658 & 94.71 & 19.718 & 110 & 26.7 & 5.9 & 4.134 & I & 25 & 1.2 & 0.019 & 4152 \\
EN031217\_171520 & 77.89 & 13.324 & 4.6 & 21.3 & 3.1 & 0.865 & I & 19.4 & 0.59 & 0.020 & 4813 \\
EN201217\_223459 & 90.86 & 21.619 & 0.92 & 36.5 & 7.4 & 0.001 & I & 44.8 & 2.8 & 0.036 & 4741 \\
EN180118\_182623 & 87.62 & 20.056 & 3 & 26.8 & 6.0 & 0.135 & I & 26.2 & 1.14 & 0.017 & 5223 \\
EN280218\_031712 & 77.65 & 14.572 & 0.039 & 39.2 & 7.1 & 0.003 & I & 104.5 & 1.12 & 0.032 & 5212 \\
EN010318\_025150 & 77.92 & 14.253 & 3.4 & 34.7 & 8.1 & 0.004 & II & 23.5 & 3.32 & 0.098 & 4814 \\
EN010318\_041356 & 78.96 & 16.841 & 0.076 & 35.3 & 6.8 & 0.008 & I & 79.3 & 0.9 & 0.019 & 4965 \\
EN180318\_190027 & 82.43 & 13.605 & 2.5 & 27.7 & 4.6 & 0.103 & I & 26.7 & 1.2 & 0.039 & 4430 \\
EN080418\_184736 & 88.65 & 16.415 & 61 & 25.2 & 2.3 & 0.119 & I & 16.1 & 2.12 & 0.047 & 4191 \\
EN100518\_214158 & 74.60 & 14.150 & 0.16 & 44.8 & 8.7 & 0.002 & II & 119 & 2.35 & 0.070 & 3140 \\
EN120518\_204142 & 73.50 & 14.122 & 1.8 & 39.0 & 8.7 & 0.004 & II & 41.2 & 3.28 & 0.099 & 2433 \\
EN230518\_194647 & 80.37 & 12.898 & 7.6 & 23.7 & 3.2 & 0.204 & I & 17.3 & 1.29 & 0.046 & 5399 \\
EN260518\_204814 & 86.81 & 13.978 & 1.4 & 35.2 & 7.5 & 0.081 & I/II & 49.9 & 1.33 & 0.041 & 4015 \\
EN290518\_012837 & 72.69 & 12.034 & 0.23 & 41.0 & 8.8 & 0.069 & I/II & 77.5 & 0.86 & 0.036 & 3838 \\
EN150618\_200523 & 86.15 & 15.457 & 5.3 & 26.8 & 5.7 & 0.395 & I & 24.6 & 1 & 0.025 & 3978 \\
EN300618\_013421 & 84.83 & 18.558 & 17 & 34.4 & 9.2 & 0.012 & II & 25 & 3.21 & 0.056 & 3259 \\
EN300618\_210658 & 85.89 & 24.815 & 2.2 & 35.7 & 8.5 & 0.006 & I & 49.9 & 2.23 & 0.022 & 3089 \\
EN110918\_015402 & 78.30 & 15.676 & 0.27 & 44.6 & 7.2 & 0.002 & I & 159.2 & 2.07 & 0.051 & 3510 \\
EN110918\_214648 & 91.43 & 23.651 & 9.2 & 26.1 & 7.2 & 0.081 & I & 17.3 & 1.74 & 0.019 & 4159 \\
EN170918\_185913 & 71.42 & 13.595 & 0.011 & 47.5 & 10.0 & 0.002 & I/II & 173.1 & 1.24 & 0.040 & 3453 \\
EN081018\_195513 & 82.05 & 13.978 & 3 & 23.6 & 3.2 & 0.567 & I & 27.1 & 0.59 & 0.018 & 4071 \\
EN121018\_213150 & 80.56 & 15.758 & 0.38 & 36.3 & 9.9 & 0.022 & I & 44.4 & 1.57 & 0.038 & 5233 \\
EN071118\_010142 & 66.46 & 13.645 & 12 & 37.4 & 9.3 & 0.126 & II & 33.3 & 2.84 & 0.091 & 4287 \\
EN161118\_172527 & 78.31 & 14.195 & 2.8 & 33.7 & 8.0 & 0.015 & II & 25.7 & 2.55 & 0.076 & 3836 \\
EN291118\_041019 & 90.73 & 25.820 & 11 & 22.5 & 3.5 & 0.424 & I & 21 & 1.11 & 0.010 & 4334 \\
 \hline
 \end{tabular}
 }
 \end{table}

Finally, analysis of suspected iron meteoroids  \cite{Vojavcek2020PSS, Borovicka2022AA_II} revealed bulk densities and ablation coefficients that do not align with typical iron properties \cite{Gritsevich2012CosRe, Turchak2014JTAM, Kyrylenko2023ApJ}. Instead, our ablation coefficients were significantly higher (0.25), and only 1 out of 12 best fit densities was close to be compatible in density ($\sim$6000 $kg\,m^{-3}$).

In ten events in the EN catalog, the terminal velocities exceed the entry velocities. A slight negative discrepancy is observed in certain cases, attributed to instances where the velocity error is significant. This difference is consistently much smaller than the associated error. We also note that two events had zero photometric mass but non-zero terminal mass. Photometric mass and total radiated energy were unavailable for these fireballs due to the absence of photometric data. Both values are recorded as zero, whereas dynamic data were available. In this regard, it should be mentioned that our results are predicated on the assumption that beginning photometric and terminal dynamic mass are correlated and represent the real properties of the meteoroid. Small changes in the mass constraints can significantly influence the results, particularly the retrieved bulk density. In instances where this does not happen, our inferences do not provide parameters with physical significance. Instead, it is retrieved values that reconcile the available data to the greatest possible extent. 

Overall, the observed decline in model accuracy when applying the $P_E$ classification to fireball events serves as a critical indicator of the constraint's limiting effects on the purely dynamical model's predictive capabilities. One interpretation could be that the $P_E$ criterion, while useful for classification, may not be universally applicable for detailed dynamic modeling of fireball events, underscoring the need for a more adaptable modeling framework. Another possibility could be related to the multi-step complex fireball reduction process employed for the EN catalog or to the inability of our model to represent complex fragmentations or/and non-decelerating fireballs. Also, photometric mass is likely better constrained for $P_E$ type I events due to the availability of numerous meteorite falls for recalibration and the significant refinement of initial luminous efficiency estimates. In contrast, fragile and fast impactors present greater uncertainty, which may influence our fitted parameters. In future efforts, we will explore whether the $\alpha$-$\beta$ model can overcome these challenges when information about an intermediate point in the flight is incorporated into the optimization process.


\section{Conclusions} \label{sec:conclusions}

In this study, we have introduced a methodology to infer the velocity profile parameter space of fireballs and their corresponding meteoroid bulk densities, ballistic coefficients, mass loss rates, and ablation coefficients from incomplete datasets. For instance, these datasets can lack full observational records of the atmospheric flight of fireballs. By employing reverse engineering paradigms, we have utilized metaheuristic global optimization algorithms, specifically differential evolution algorithms. These computational techniques, in conjunction with theoretical models of atmospheric entry predicated on single-body dynamics, have facilitated the derivation of coherent estimations for the parameters associated with fireballs. The parameters were retrieved from the EN fireball catalog, contingent upon the initial and terminal heights and velocities, as well as the initial and terminal masses used as constraints. This task presents inherent challenges due to the necessity to combine two disparate methodologies: the initial mass deduced from photometric data and the terminal mass computed through dynamical analysis, both integrated in this study within the framework of the purely dynamical $\alpha$-$\beta$ model.

Our approach has involved performing optimizations under two distinct scenarios: one that imposes bounds on the bulk density and ablation coefficient parameters based on the $P_E$ criterion classification, and another devoid of any such constraints. When the $P_E$ constraint was implemented, we have observed a tangible decrease in the proportion of events that satisfied the conditions of a terminal height residual less than 100 $m$, velocities residuals smaller than 100 $m\,s^{-1}$, and masses residuals bellow 1 $g$.

We have validated the retrieved fireball parameters primarily for $P_E$ type I or $Pf$-I and $Pf$-II events by ensuring low fit residuals and comparing the general performance in terminal height predictions against parameters derived using complete point-by-point observations. This evaluation is further supported by the five EN examples available to us with full trajectory measurements. These examples demonstrate that the velocity profile aligns well with observations for a $Pf$-I event and remains acceptable for the $Pf$-II case. However, we acknowledge that even with low fit residuals, our proposed implementation approach of $\alpha$-$\beta$ encounters challenges in accurately modeling the final flight phase, likely due to limitations in representing fragmentation processes along the trajectory based solely on the initial and terminal points constrains. In general, objects with asteroidal compositions were consistently well-fitted when no $P_E$ constrains were applied. Conversely, the least accurate results occurred for fragile high-velocity meteoroids, especially in scenarios where deceleration approached nullity. In addition, this new approach did not allow us to confirm most of the impactors previously flagged as iron meteoroids. This underscores the need for further refinement of the method to better account for more complex dynamics. We also acknowledge the need for further validation using complete atmospheric flight data from a broader sample of fireballs.


A noteworthy observation is that the $\alpha$ and $\beta$ values derived from our analysis are significantly different relative to those obtained from other databases where only dynamical models were applied (without constraining by photometric measurements). This disparity may arise from the incorporation of initial and/or terminal mass constraints in our fitting process, differing from previous works and influencing the calculated mass loss rate required to ensure the matching fit. We have also reconstructed the $\alpha$-$\beta$ diagram for the EN catalog, presenting it with and without the inclusion of the slope dependency. The diagram exhibited a degree of clustering not observed in other databases, potentially indicating either particularity inherent to the data reduction process, a fundamental incompatibility between the photometric and dynamical methodologies in a majority of cases, or incorrect mass determinations. 


Our key findings are summarized as follows:
\begin{itemize}

    \item The use of metaheuristic optimization and the $\alpha$-$\beta$ method provides an effective framework for constraining atmospheric flight parameters of asteroid-like meteoroids even from incomplete datasets, particularly those lacking point-by-point observations.
    
    \item When using the $\alpha$-$\beta$ method to reconstruct velocity profiles based on two points from datasets similar to the EN catalog, avoiding $P_E$ classification constraints enhances accuracy.
    
    \item 44\% of the EN catalog was fitted under $P_E$ restrictions, satisfying the residual conditions.
    
    \item 90\% of the EN catalog was fitted without restrictions, satisfying the residual conditions.
    
    \item 26\% of the EN catalog was fitted without restrictions, satisfying the residual conditions and aligning with the previous $P_E$ classification based on the retrieved bulk densities and ablation coefficients.

    \item Despite achieving low fit residuals in some cases, the $\alpha$-$\beta$ approach implemented here does not accurately constrain the final flight deceleration of meteoroids, particularly for fragile high-velocity ones. This limitation likely arises from challenges in representing complex fragmentation processes or atmospheric flights with negligible deceleration. Alternatively, it may result from the imposed mass-loss rate needed to reconcile the photometric mass at entry with the estimated terminal mass, which is constrained solely by the observed deceleration.

\end{itemize}

Our findings offer a new perspective on the utility of trajectory constraints and data reduction techniques employed in fireball analysis. The methodology we have introduced is especially pertinent when dealing with inaccessible datasets, such as older fireball observations, or when information is unobtainable. By exploring these novel methods, we aim to overcome the constraints associated with conventional data-reduction techniques that rely on point-by-point observational data. Our objective is to forge ahead in understanding the properties of meteoroids entering the Earth's atmosphere. Additionally, we aim to contribute to creating robust, yet more adaptable tools for characterizing fireball events. Such advancements will not only augment our capacity to recover fresh meteorites but will also enhance our preparedness to monitor and mitigate the hazards posed by near-Earth objects.


\acknowledgments
This work was supported by the LUMIO project funded by the Agenzia Spaziale Italiana (2024-6-HH.0). We express gratitude to the Finnish Geospatial Research Institute and the Academy of Finland for supporting the project no. 325806 (PlanetS), which facilitated the development of the analytical methods presented in this paper. The program of development within Priority-2030 is acknowledged for supporting the research at UrFU. EP-A has carried out this work in the framework of the project Fundación Seneca (22069/PI/22), Spain. We thank Jiří Borovička for sharing point-by-point measurements for five EN fireballs.

\section*{Open Research}

The fireball catalog used in this work is publicly available in \citeA{Borovicka2022AA_I, Borovicka2022AA_II}. The computational tool developed for inferring the fireball velocity profiles and the characteristic parameters can be accessed via Zenodo \cite{Eloy_2024_inferring_source}. Figures were made with the Matplotlib library \cite{Hunter2007}.

\bibliography{agusample}

\clearpage
\section{Appendix}
\appendix
\section{Summary of definitions and abbreviations}\label{sec:defs}
\begin{table}[H]
\begin{tabular}{lll}
$A$	&	 $-$ 	&	Shape factor, a cross sectional area to volume ratio $A=S\left(\frac{\rho_m}{m}\right)^{2/3}$.	\\
$c_d $ 	&	 $-$ 	&	 Drag coefficient.	\\
$c_h $ 	&	 $-$ 	&	 Heat-transfer coefficient.	\\
$\overline{Ei}$&	 $-$ 	&	 Exponential integral, $\overline{Ei}(x)=\int_{\infty}^{x}\frac{e^{z}}{z}\,dz\,$.\\
$h$ 	&	 $-$ 	&	 Meteoroid height above sea level ($m$).	\\
$h_{beg}$ 	&	 $-$ 	&	 Meteoroid height above sea level at the beginning of the luminous trajectory ($m$).	\\
$h_{ter}$ 	&	 $-$ 	&	 Meteoroid height above sea level at the terminal point of the luminous trajectory ($m$).	\\
$h_0$ 	&	 $-$ 	&	 Scale height of the homogeneous atmosphere ($h_0=7.16\,km$).	\\
$H^*$ 	&	 $-$ 	&	 Effective destruction enthalpy ($J\, kg^{-1}$).	\\
$m$ 	&	 $-$ 	&	 Normalized meteoroid mass, $m = \cfrac{M}{M_{beg}}$  (dimensionless).	\\
$M$ 	&	 $-$ 	&	 Meteoroid mass ($kg$).	\\
$M_{beg}$ 	&	 $-$ 	&	 Initial entry mass of meteoroid at the beginning of the luminous trajectory ($kg$).	\\
$M_{ter}$ 	&	 $-$ 	&	 Terminal mass of meteoroid at the end of the luminous trajectory ($kg$).	\\
$S$ 	&	 $-$ 	&	 Cross sectional area of the body ($m^2$).	\\
$v$ 	&	 $-$ 	&	 Normalized meteoroid velocity, $v = \cfrac{V}{V_{beg}}$  (dimensionless).	\\
$V$ 	&	 $-$ 	&	 Meteoroid velocity ($km\, s^{-1}$).	\\
$V_{beg}$ 	&	 $-$ 	&	 Initial entry velocity of the meteoroid at the beginning of the luminous trajectory ($km\, s^{-1}$).	\\
$V_{ter}$ 	&	 $-$ 	&	 Terminal velocity of the meteoroid at the end of the luminous trajectory ($km\, s^{-1}$).	\\
$y$ 	&	 $-$ 	&	 Normalized meteoroid height, $y = \cfrac{h}{h_0}$  (dimensionless).	\\
$y_{ter}$ 	&	 $-$ 	&	 Normalized meteoroid terminal height, $y_{ter} = \cfrac{h_{ter}}{h_0}$  (dimensionless).	\\
$\alpha$	&	 $-$ 	&	Ballistic coefficient.\\
$\beta$	&	 $-$ 	&	Mass loss parameter.	\\
$\gamma $ 	&	 $-$ 	&	Angle of the meteoroid flight to the horizontal.	\\
$z$ 	&	 $-$ 	&	Angle of the meteoroid flight to the vertical.	\\
$\mu$	&	 $-$ 	&	Shape change coefficient representing the rotation of the meteoroid ($0\leq\mu\leq2/3$).	\\
$\rho$ 	&	 $-$ 	&	Atmospheric density ($kg\, m^{-3}$).	\\
$\rho_{beg}$ 	&	 $-$ 	& Atmospheric density at the beginning of the observed luminous phase ($g\, cm^{-3}$).	\\
$\rho_{sl}$ 	&	 $-$ 	&	Atmospheric density at sea level ($\rho_{sl}=  1.29 kg\, m^{-3}$).	\\
$\rho_m$ 	&	 $-$ 	&	Meteoroid bulk density ($kg\,m^{-3}$).	\\
$\sigma$	&	 $-$ 	&	Ablation coefficient, $\sigma = \cfrac{c_h}{H^*c_d}\quad$ ($s^2\,km^{-2}$).	\\
$\tau$ 	&	 $-$ 	&	Luminous efficiency coefficient.	\\
$\mathfrak{r}$ & $-$ & Residuals \\
$k_B$ & $-$ & Criterion for fireball classification proposed by \citeA{Ceplecha1967SCoA}.\\
$P_E$ & $-$ & Criterion for fireball classification proposed by \citeA{Ceplecha1976JGR}.\\
$ln(2\alpha\beta)$ & $-$ & Criterion for fireball classification proposed by \citeA{moreno2020physically}.\\
$Pf$ & $-$ & Criterion for fireball classification proposed by \citeA{Borovicka2022AA_II}.\\
$h_I$ & $-$ & Formula for predicting terminal height \cite{Moreno2015Icar}. \\
$h_{II}$ & $-$ & Formula for predicting terminal height \cite{Moreno2015Icar}. \\
$h_{III}$ & $-$ & Formula for predicting terminal height \cite{Moreno2015Icar}. \\
\end{tabular}
\end{table}

\section{Dimensional Analysis} \label{sec:dim}
The Buckingham $\pi$ theorem is a fundamental theorem of dimensional analysis. The theorem states that if there is a relationship among n physical quantities, which does not change its form when the units of measurement are scaled within a certain class of unit systems, then it is equivalent to a relationship among, generally, a smaller number  $p$ = $p_1$ - $p_2$ of dimensionless quantities, where $p_2$ is the number of physical dimensions involved among the original $p_1$  quantities.  

The $\pi$ theorem implies that the validity of the physical laws does not and should not depend on a specific unit system, and any physical law can be expressed by involving only dimensionless combinations of the variables linked by the law. Consequently, by introducing fewer dimensionless parameters (in our case two, $\alpha$ and $\beta$) in the model compared to all the unknowns they entail, a unique yet self-similar solution can be found that is characteristic for any given fireball event. This approach, besides being physically sound, offers a unique interpretation for every fireball event, avoids the necessity to assign random values to an ``excess'' number of unknowns, and is not prone to overfitting, ensuring robust performance on new, untrained datasets. This efficiency in data utilization is complemented by the insight generation capability of physical models, offering valuable understanding into underlying processes and mechanisms.

\section{Exponential integral}\label{sec:exp_int}
Generally speaking, the exponential integral $Ei$ is a special function on the complex plane.

For real non-zero values of $x$, the exponential integral $Ei(x)$ is defined as
\begin{equation}
    Ei(x) = -\int_{-x}^{\infty} \frac{e^{-t}}{t} \,dt = \int_{-\infty}^{x} \frac{e^{t}}{t} \,dt.
\end{equation}

The Risch algorithm shows that $Ei$ is not an elementary function. The definition above can be used for positive values of $x$ (hence, for the whole range of physical values of the arguments in meteor physics equations), and the integral has to be understood in terms of the Cauchy principal due to the singularity of the integrand at zero. Practical approximations for $y(v)$ using elementary functions have also been developed, as discussed in \citeA{Gritsevich2016}.  

The Taylor series expansion of the exponential integral $Ei(x)$ around $x = 0$ can be practical in calculations and is given by:
\begin{equation}
    Ei(x) = \gamma + \ln |x| + \sum_{n=1}^{\infty} \frac{x^n}{n \cdot n!}
\end{equation}
where $\gamma$ is a mathematical constant called the Euler's constant (or the Euler–Mascheroni constant), defined as the limiting difference between the harmonic series and the natural logarithm, represented here by $\log$:

\[
\gamma = \lim_{n \to \infty} \left( -\ln n + \sum_{k=1}^{n} \frac{1}{k} \right) = \int_{1}^{\infty} \left( -\frac{1}{x} + \frac{1}{\lfloor x \rfloor} \right) \, dx.
\]

Here, $\lfloor \cdot \rfloor$ denotes the floor function.

The numerical value of Euler's constant, to 5 decimal places, is approximately:

\[ \gamma \approx 0.57721. \]

A faster converging series for the exponential integral $Ei$ was found by Srinivasa Ramanujan:
\begin{equation}
    Ei(x) = \gamma + \ln x + e^{\left(\frac{x}{2}\right)} \sum_{n=1}^{\infty} \frac{(-1)^{n-1} x^n}{n! \, 2^{n-1}} \sum_{k=0}^{\lfloor (n-1)/2 \rfloor} \frac{1}{2k+1}
\end{equation}

\end{document}